\documentclass[conference]{IEEEtran}
\IEEEoverridecommandlockouts
\usepackage{cite}
\usepackage{amsmath,amssymb,amsfonts}
\usepackage{algorithmic}
\usepackage{graphicx}
\usepackage{textcomp}
\usepackage{xcolor}
\def\BibTeX{{\rm B\kern-.05em{\sc i\kern-.025em b}\kern-.08em
    T\kern-.1667em\lower.7ex\hbox{E}\kern-.125emX}}

\usepackage[nowarn,acronyms,nonumberlist,nopostdot,nomain,nogroupskip]{glossaries}

\usepackage{caption}
\usepackage{subcaption}

\newacronym{lidar}{LIDAR}{Light Detection and Ranging}
\newacronym{rf}{RF}{Radio Frequency}
\newacronym{mimo}{MIMO}{Multiple-Input-Multiple-Output}
\newacronym{bs}{BS}{Base Station}
\newacronym{lis}{LIS}{Large Intelligent Surface}
\newacronym{mf}{MF}{Matched Filter}
\newacronym{nn}{NN}{Neural Network}
\newacronym{snr}{SNR}{Signal-to-Noise Ratio}
\newacronym{fmcw}{FMCW}{Frequency-Modulated Continuous-Wave}
\newacronym{adc}{ADC}{Analog to Digital Converter}
\newacronym{fft}{FFT}{Fast Fourier Transform}
\newacronym{cfar}{CFAR}{Constant False Alarm Rate}
\newacronym{gnn}{GNN}{Graph Neural Network}
\newacronym{knn}{KNN}{K-Nearest Neighbor}
\newacronym{mpnn}{MPNN}{Message Passing Neural Network}
\newacronym{mlp}{MLP}{Multi Layer Perceptron}
\newacronym{mse}{MSE}{Mean Squared Error}
\newacronym{iou}{IOU}{Intersection Over Union}
\newacronym{los}{LOS}{Line-of-Sight}

\usepackage{pgfplots}
\usepackage{svg}  
\usetikzlibrary{plotmarks}
\usetikzlibrary{arrows.meta}
\usepgfplotslibrary{patchplots}

\usepackage{svg}

\pgfplotsset{compat=1.17}

\begin{document}
\title{User Localization using RF Sensing: A Performance comparison between LIS and mmWave Radars
\thanks{This project has received funding from the European Union’s Horizon 2020 research and innovation programme under the Marie Skłodowska-Curie Grant agreement No. 813999.}
}

\author{
Cristian J. Vaca-Rubio$^{1,*}$\thanks{$^*$Both authors contributed equally to this research.}, Dariush Salami$^{2,*}$, Petar Popovski$^1$, \\ Elisabeth de Carvalho$^1$, Zheng-Hua Tan$^1$ and Stephan Sigg$^2$ \\
$^1$Department of Electronic Systems, Aalborg University, Aalborg, Denmark \\
$^2$Department of Communications and Networking, Aalto University, Espoo, Finland \\
\{cjvr, petarp, edc, zt\}@es.aau.dk, \{dariush.salami, stephan.sigg\}@aalto.fi \\

}

\maketitle

\begin{abstract}
Since electromagnetic signals are omnipresent, \gls{rf}-sensing has the potential to become a universal sensing mechanism with applications in localization, smart-home, retail, gesture recognition, intrusion detection, etc. 
Two emerging technologies in \gls{rf}-sensing, namely sensing through \glspl{lis} and mmWave \gls{fmcw} radars, have been successfully applied to a wide range of applications. 
%Accurate localization of targets is of a high importance especially in industrial environments where robots and humans work together and safety cannot be compromised. Although there exist approaches to tackle the problem using both \gls{lis} and mmWave radars in the literature, there is still a need for comparison between these two sensing technologies. 
In this work, we compare \gls{lis} and mmWave radars for localization in real-world and simulated environments.  
%First, a set of comprehensive experiments is designed in a real-world environment to localise targets using an mmWave radar. 
%Then, the same set of experiments is conducted using an \gls{lis} in a simulation environment on the sub-mmWave band. 
%MmWave radars transmit a sinusoid with an increasing frequency through time (chirp) and sense the target processing the reflected signal, while in \gls{lis} we can sense the target in two ways: assuming the target as an active transmitter or taking advantage of commodity transmitters to sense passive target from their signal reflections. 
%We compare the accuracy of localization in both cases. 
In our experiments, the mmWave radar achieves 0.71 \gls{iou} and 3cm error for bounding boxes, while \gls{lis} has 0.56 \gls{iou} and 10cm distance error. 
Although the radar outperforms the \gls{lis} in terms of accuracy, \gls{lis} features additional applications in communication in addition to sensing scenarios.
\end{abstract}

\glsresetall

\section{Introduction}
Recently, with the advance of robots, remote controlled and autonomous vehicles (UAVs, vehicles), localization and positioning has become a key support capability. 
%Localization accounts for estimating the real-world position of the target in the environment. 
RGB cameras~\cite{kononov2022external}, lidars~\cite{zhang2022efficient}, and \gls{rf} sensors~\cite{palipana2017recent} are common approaches used for localization. 
Lidars are capable of providing an accurate 3D model of an environment. 
Their performance is affected by weather conditions though~\cite{li2020millimeter}. 
This is similar for RGB cameras, which also have a reduced performance in challenging lighting conditions~\cite{palipana2021pantomime}. 
\gls{rf}-based sensors and \gls{lis} are resilient to lighting and weather conditions\cite{palipana2021pantomime}.

\begin{figure}[t]
    \centering
    \includegraphics[width=0.9\columnwidth]{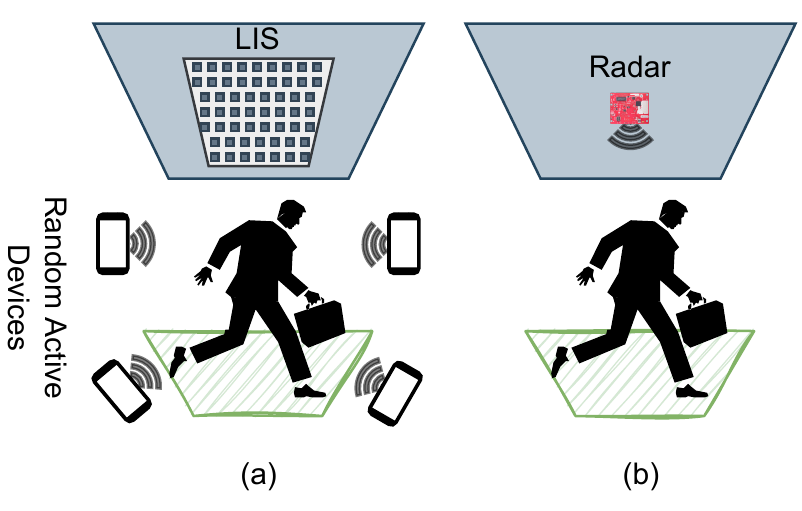}
    \caption{Schematic comparison of the two systems (a) LIS-based, (b) radar-based}
   \label{fig:schematic}
\end{figure}

Wireless sensing is becoming integrated into future communication systems~\cite{latva2019key}. 
%Because of the growth of the \gls{mimo} technology~\cite{larsson2014massive}, there has been a lot of study on improving communication systems. 
%\glspl{bs} are developed with an even larger number of antennas than in massive \gls{mimo}. 
Specifically, with the increasing count of antennas added to \glspl{bs}, researchers have exploited first \gls{mimo}, then massive \gls{mimo}, and now \gls{lis}, a continuous electromagnetic surface, comprising a large number of tightly spaced small antenna elements, capable of transmitting and receiving electromagnetic signals. 
%In actuality, they are planar arrays made up of a large number of tightly spaced small antenna components. 
\gls{lis} are capable to support communication as well as sensing~\cite{vaca2020assessing}.
Likewise, mmWave radars have been widely used in many sensing applications from gesture recognition~\cite{salami2021integrating,salami2022tesla,salami2021zero} to localization~\cite{palipana2017recent}. 
%given their capability of sensing fine-grained movements, robustness to environmental lighting or weather conditions, and penetration through thin, non-metallic surfaces. 

In this work, we compare, for the first time in the literature, these two technologies to solve a sensing-based localization task as it is shown in Fig.~\ref{fig:schematic}. 
We perform experimental measurements using an \gls{fmcw} radar, namely IWR1443, operating in
the 77-81 GHz RF-band. 
Since no \gls{lis} prototypes exist yet in the market, we perform a comprehensive simulation based on ray-tracing, mimicking the experimental environment. 
In particular, we will reconstruct the radio map of the propagation environment using an indoor \gls{lis} deployment in the ceiling, as also described in~\cite{vaca2021radio}. 
%We will use this method to compare both solutions in a passive user localization problem.

\section{System Models and Problem Formulation}
In this section, we discuss the configurations of the two systems. 
The \gls{lis} system is implemented in a simulation environment, while the radar system has a real-world test-bed.

\subsection{\gls{lis} System Model}
We consider an indoor setting in which $D_a$ active devices are present. 
The devices transmit for communication. 
Sensing with the \gls{lis} is then subsequently performed by piggybacking these signals.
  
A \gls{lis} of $A$ antenna elements is placed at the room's ceiling. 
We assume an ideal \gls{lis} made up of isotropic antennas.
Therefore, physical effects such as mutual coupling are neglected.
The sensing task is to create a radio map that captures a person in the environment by superimposing the received signals from each of the $d \in D_a$ elements at each of the $A$ \gls{lis} components.
The superposed complex baseband signal received at the \gls{lis} is given by 
\begin{equation}
    \label{eq:RecSignal}
	\mathbf{y} = \sum_{d=1}^{D_a}\mathbf{h_d}x_d + \mathbf{n}.
\end{equation}
Here, $x_d$ is the transmitted symbol from device $d$ (without loss of generality, we consider $x_d=1$), $\mathbf{h_d}\in\mathbb{C}^{N\times 1}$ is the channel vector from a specific position of device $d$ to each antenna-element, and $\mathbf{n}\sim\mathcal{CN}_{N}(\mathbf{0},\sigma^2\mathbf{I}_N)$ is the noise vector. 
For simplicity, we consider narowband transmissions and this removes the effect of frequency selectivity.

\begin{figure}[t]
    \centering
    \includegraphics[width=0.9\columnwidth]{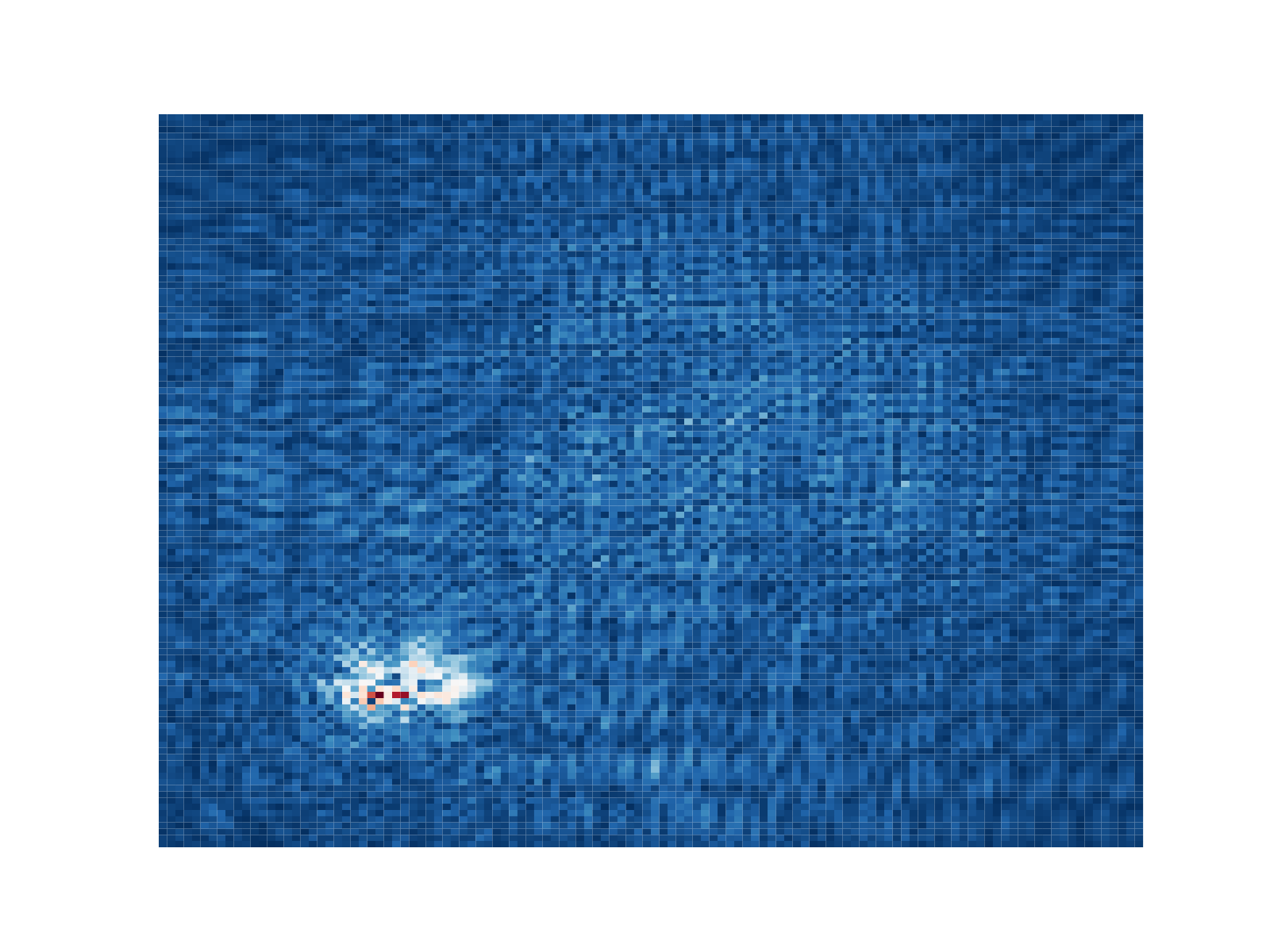}
    \caption{A radio map obtained for an $M=118\times118$ antenna elements $\frac{\lambda}{2}$ spaced \gls{lis} in a noiseless scenario with $D_a=5$ active devices by using a \gls{mf} design for $f=3.5$ GHz, $d=3.2$ m and $N_f=100\times100$ antenna elements $\frac{\lambda}{2}$ spaced.}
   \label{fig:exemplary_map}
\end{figure}

The large physical dimensions of the \gls{lis}, compared with the distance from the transmitters to the ceiling, lead to a spherical-wave propagation condition. 
The spherical-wave channel coefficient $h_{sp,a}$ at the $a$-th element from an arbitrary active device transmission is proportional to~\cite{zhou2015spherical}
\begin{equation} h_{sp,a} \ \propto\ \frac{1}{d_a}e^{-j\frac{2\pi}{\lambda}d_a},\label{eq:pattern} \end{equation}
where $d_a$ denotes the distance between the active device and the $a$-th antenna. 
To compute the radio map, we can derive a \gls{mf} such that~\cite{vaca2021radio}
\begin{equation}\label{eq:r_map} \mathbf{y_{mf}} =  \mathbf{h_{sp}} \ast \mathbf{y}. \end{equation}
In eq~\ref{eq:r_map}, $\ast$ denotes the convolution operator, $\mathbf{h_{sp}} \in\mathbb{C}^{N_f\times 1}$ is the expected spherical pattern (steering vector) for $N_f$ antennas \gls{lis} deployment on (\ref{eq:pattern}), $\mathbf{y}$ is the received signal from (\ref{eq:RecSignal}) and $\mathbf{y_{mf}} \in\mathbb{C}^{A\times 1}$ is the filtered output that represents the radio map. 
We zero-pad $\mathbf{y}$ such that we guarantee $\mathbf{y_{mf}} \in\mathbb{C}^{A\times 1}$ dimension to perform the convolution. 
To obtain the radio map, we compute the energy at the output of the \gls{mf}. 
In this way, we are measuring the energy of the signals reflected from the target. 
In order to design the filter, we assume knowledge of the frequency $f$ and the distance $d$ (the $z$ coordinate, since the \gls{lis} is deployed at the ceiling) between the transmitter and the \gls{lis}\footnote{A detailed explanation of the radio maps is given in~\cite{vaca2021radio}. 
The distance $d$ is a parameter for the filter design and may differ from the actual distance to the transmitter in the evaluation.}. 
Figure~\ref{fig:exemplary_map} shows the detection of the target on the \gls{lis} processed radio map.
\subsubsection{Simulated scenario}
\label{subsubsec:Scenario}
 We perform simulations via ray tracing~\cite{FEKO} in a $ 4.7\times 4.7 \times 3.2$ m simulated area. 
 We deploy a \gls{lis} with $118\times118$ elements separated by $\lambda/2$. 
 The $D_a$ active devices transmit narrowband signals of 20 dBm at 3.5 GHz and they are randomly deployed in the space $(x, y, z)$. 
 The distance from which the \gls{mf} is calibrated is $d=3.2$, as we set a kernel for the procedure such that it scans the entire height of the room. 
 The human is modeled as a rectangle of dimensions 0.3x0.5x1.83 m (reasonable human dimensions obtained from~\cite{potkany2018requirements}) with conductivity $s=1.44$ S/m, relative permittivity $\varepsilon=38.1$ and relative permeability $\mu=1$~\cite{hall2007antennas}. 
 To be fair with the experimental setup, the positions of the passive human $(x, y, z=1.83)$ m are randomly selected in a circle of radius 2.5 m.
 
\subsubsection{Received signal and noise modeling}
From the ray-tracing simulation, the received signal in (\ref{eq:RecSignal}) is obtained as the complex electric field arriving at the $n$-th antenna element, $\widetilde{E}_{a}$, which can be regarded as the superposition of each ray path $r \in N_r$ from every $d \in D_a$ device, i.e., 
\begin{equation}
    \label{eq:Esum}
  \widetilde{E}_{a} = \sum_{d=1}^{D_a}\sum_{r=1}^{N_r} \widetilde{E}_{a,r,d}= \sum_{d=1}^{D_a}\sum_{r=1}^{N_r}E_{a,r,d} e^{j\phi_{a,r,d}}.  
\end{equation}
The complex signal at the output of the $n$-th element is therefore given by 
\begin{equation}
    \label{eq:complexSignal}
    y_a = \sqrt{\frac{\lambda^2Z_a}{4\pi Z_0}} \widetilde{E}_{a} + n_a.
\end{equation}
Here, $\lambda$ is the wavelength, $Z_0 = 120\pi$ os the free space impedance and  $Z_n$ is the antenna impedance. 
For simplicity, we consider $Z_n = 1\,\forall\, n$. 
Finally, we define the average \gls{snr}, $\overline{\gamma}$, as
\begin{equation}
    \label{eq:snr}
    \overline{\gamma} \triangleq  \frac{\lambda^2}{4\pi Z_0 M \sigma^2}\displaystyle\sum_{a=1}^{A} |\widetilde{E}_{a}|^2,
\end{equation}
where $A$ denotes the number of antenna elements in the \gls{lis}.

\subsection{Radar System Model}
\begin{figure*}
    \centering
    \includegraphics[width=1\linewidth]{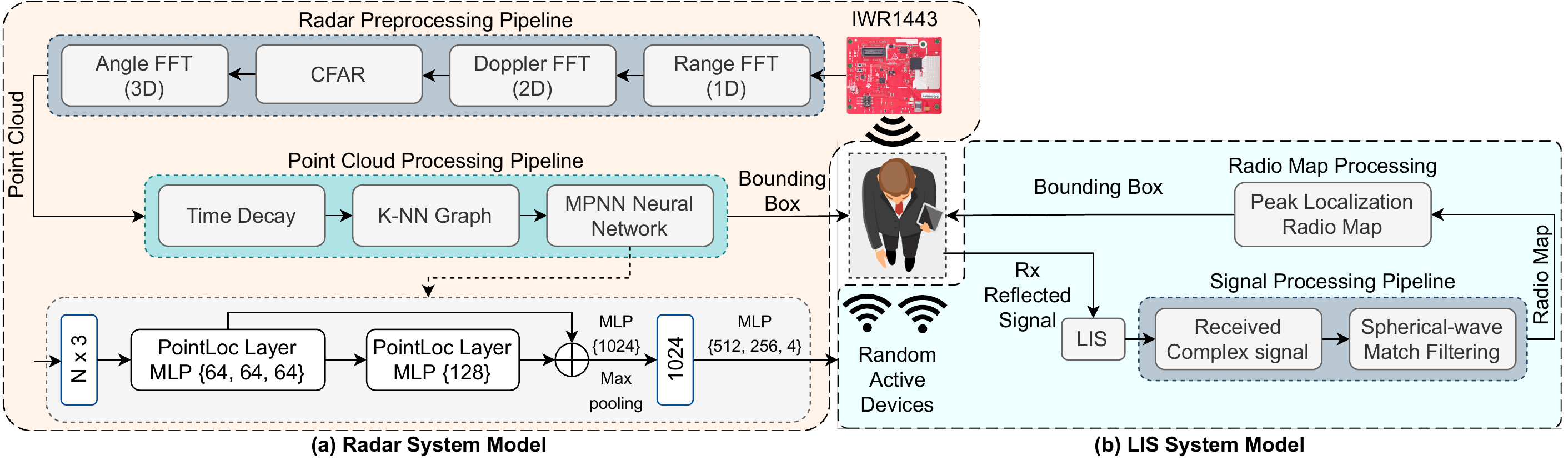}
    \caption{(a) Overview structure of the radar-based localization system. The  radar transforms the IQ samples into a point cloud through the radar processing pipeline. Then, the point cloud processing pipeline predicts a bounding box using as the localization output. (b) Overview structure of the LIS-based localization system. The \gls{lis} processes the reflected complex signal by applying the \gls{mf} procedure to obtain the radio map. Then, the radio map processing part predicts a bounding box around the target user. The random $D_a$ active devices are piggybacked to perform the sensing.}
    \label{fig:system_models}
\end{figure*}
We utilize the IWR1443 mmWave \gls{fmcw} radar operating at 77GHz with 4GHz bandwidth. 
%A four step processing unit on the evaluation kit of the radar generates point cloud data (a set of points in 3D space) from an \gls{adc} output. 
We further process the point cloud data generated by the radar to perform localization.

\subsubsection{Point Cloud Generation}
The radar evaluation kit performs a four step processing pipeline to generate point cloud data from \gls{adc} data as shown in Fig.~\ref{fig:system_models}.a.

\textit{Range \gls{fft} (1D):} 
A chirp signal, i.e. a sinusoidal signal with increasing frequency through time, is generated by the radar. 
Then the mixing operation produces an intermediate frequency signal using transmitted and reflected chirps. 
The range of the system is linearly proportional to the intermediate signal's frequency, computed through \gls{fft} operations.

\textit{Doppler \gls{fft} (2D):} 
%To observe the Doppler effect, two or more time-separated chirps need to be processed. 
The radial velocity of the target is proportional to the phase difference of multiple chirps at the range-\gls{fft} peak. 
To generate a peak at the velocity of target, a \gls{fft} is applied on the signal range (i.e., a 2D-\gls{fft} or a Doppler-\gls{fft}).

\textit{\gls{cfar}:} 
For each virtual antenna, a pre-detection matrix is formed through summation of Doppler-\gls{fft} matrices. 
The \gls{cfar} algorithm~\cite{richards2010principles} then filters the pre-deteciton matrix to produce peaks corresponding to targets.

\textit{Angle-\gls{fft}:}
For each target, an angle-\gls{fft} is applied on corresponding \gls{cfar} peaks across multiple Doppler-\glspl{fft}. 
Velocity-induced phase changes are Doppler-corrected before angle-\gls{fft} calculation.

\subsubsection{Point Cloud Processing}
%Having generated point cloud using the radar preprocessing pipeline on evaluation kit, w
We follow a three-step point cloud processing tool-chain to localize the target.

\textit{Time Decay:} 
We configured the radar to sample 33 point clouds per second\footnote{we will refer to the points generated from a single chirp as frames in the following}. 
Since points in each frame are sparse ($\approx$ 5 per frame), we apply time decay to increase density. 
Empirically, a time-window of size 5 ($\approx$ 151 ms) is used to slide through time frames to increase the number of points in each frame.

\textit{Graph Generation:} 
From the time-decayed point cloud, we then build a graph. 
%to be able to process it using a \gls{gnn}-based approach.
Consider a point cloud $X=\{x_1, ..., x_n\} \subseteq \mathbb{R}^F$ where each point $x_i$ is represented by a feature set $\{f_i^1, ..., f_i^F\}$. 
A graph-\gls{knn} algorithm is exploited to produce the \gls{knn}-graph $\mathcal{G}=\{X, \mathcal{E}\}$ where $\mathcal{E} \subseteq X \times X$ is the set of directed edges between each point and its nearest neighbours in the Euclidean space. 

\textit{Graph Processing using \gls{mpnn}:} 
Following a similar approach as~\cite{wang2019dynamic}, each node's representation is updated using the proposed \gls{mpnn} layer. 
In each layer, an aggregation function is applied to the representation of the node itself and to its neighbors' features:

\begin{equation}
    \begin{aligned}
        h_{i}^0 &= x_i,\\
        h_{i}^l &=\underset{j:(i,j)\in \mathcal{E}}{\Gamma}(\Delta_\theta(h^{l-1}_i,h^{l-1}_j)),
    \end{aligned}
    \label{eq:dynamic_edge_convolution}
\end{equation}

\noindent in which, $h_i^l$ is point $i$'s representation in layer $l$, message function $\Delta_{\theta}:\mathbb{R}^F \times \mathbb{R}^F \rightarrow \mathbb{R}^{F'}$ is a learnable function with parameters $\theta$ which is implemented using \gls{mlp}, and $\Gamma$ is a channel-wise symmetric aggregation operator (e.g. $\Sigma$, max, or mean) applied on the messages of the edge emanating from each neighbor. To capture not only local but also global structures of the input point cloud, we use $\Delta_\theta(h_i, h_j)=\bar{\Delta}_\theta(h_i, h_j-h_i)$ meaning that for each edge, we concatenate the features of the incident node of the edge with the difference between the two nodes connected by the edge.

The details of the localization model is shown in Fig.~\ref{fig:system_models}. We empirically use the proposed layer in Eq.\ref{eq:dynamic_edge_convolution}, PointLoc, twice in the architecture of the model. Then we concatenate the output of the two layers followed by an \gls{mlp} to achieve a vector representation of the input point cloud with length of 1024. Finally, after applying a few layers of \gls{mlp} to extract fine-grained features, we use a linear layer with 4 neurons and without activation function to output the four dimensions of the bounding box. The loss function is \gls{mse}.
\subsection{Main Assumptions}
The main assumptions for both approaches are as follows. 
With respect to \gls{lis}, we assume that mutual coupling is ignored. 
It is commonly represented using a coupling matrix that takes into account the influence of nearby antennas~\cite{su2001modeling} but the influence can be compensated after estimation, and hence does not affect any conclusion drawn. 
%which is why we decided to neglect this effect, as it will not affect to the conclusions of this work. 
Radar and \gls{lis} have a \gls{los} to the target. 
Also, to design the filter we need to assume a-priori the frequency $f$ and the distance from a transmitter to the \gls{lis} $d$. 
However, this last parameter is not a very strong assumption (cf. Section~\ref{subsubsec:Scenario}). 
We therefore set it to be the distance as if the transmitter were on the floor. 
This means that the transmitters can be at different distances and are not limited by the filter design.

The field of view of the radar is -55$^\circ$ to +55$^\circ$. 
%So, the target should be in the field of view of the radar to be able to localized.

\section{Implementation and Evaluation}
In this section, we explain the way we collected data in the real-world environment for the mmWave radar and simulation environment for \gls{lis}. 
Then, we elaborate on the model training process, evaluation metrics, and finally the localization performance comparison between the radar and the \gls{lis}.

\subsection{Dataset}
\begin{figure}
\centering
\begin{subfigure}{.5\columnwidth}
  \centering
  \includegraphics[width=0.99\textwidth]{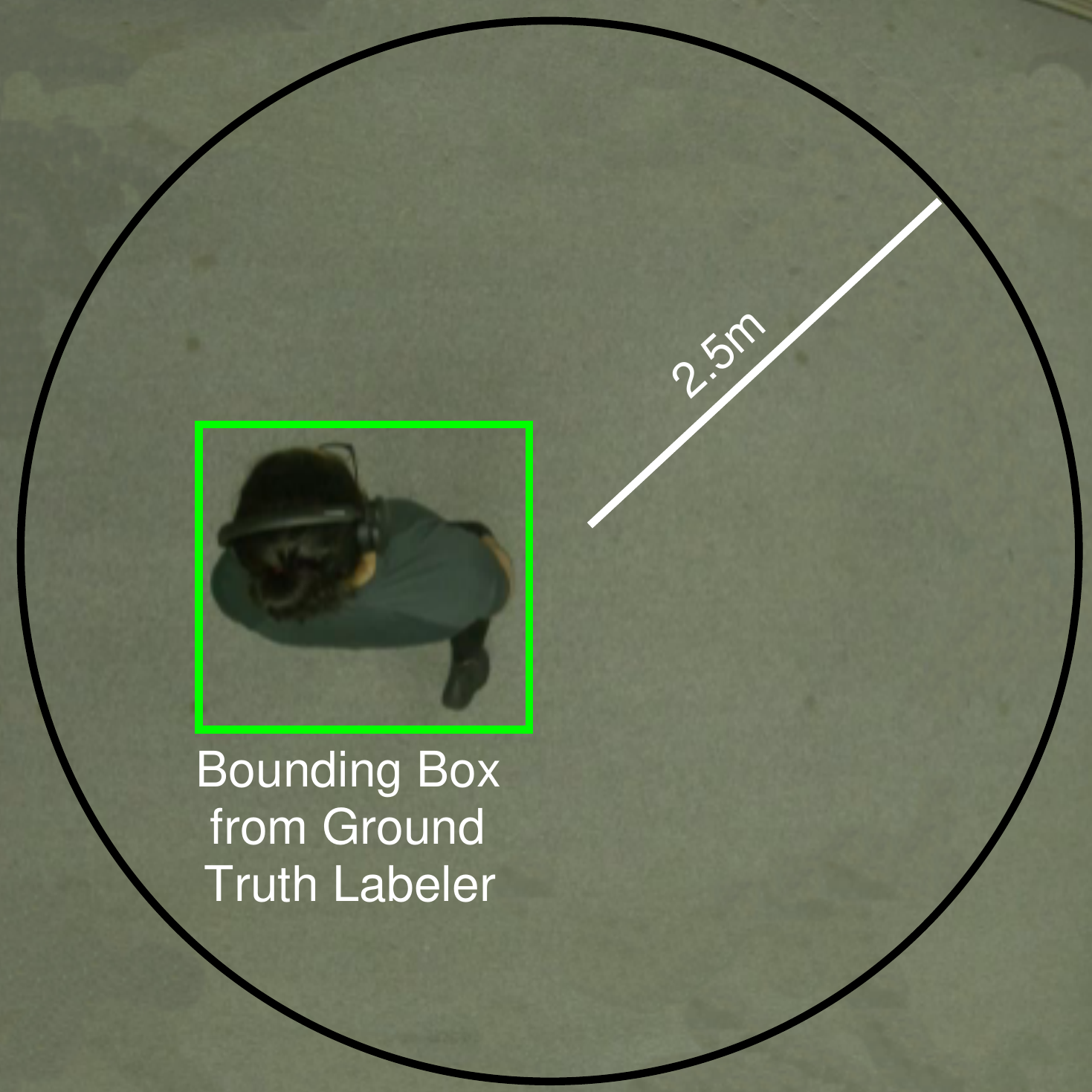}
  \caption{}
  \label{fig:radar_ground}
\end{subfigure}%
\begin{subfigure}{.5\columnwidth}
  \centering
  \includegraphics[width=0.99\textwidth]{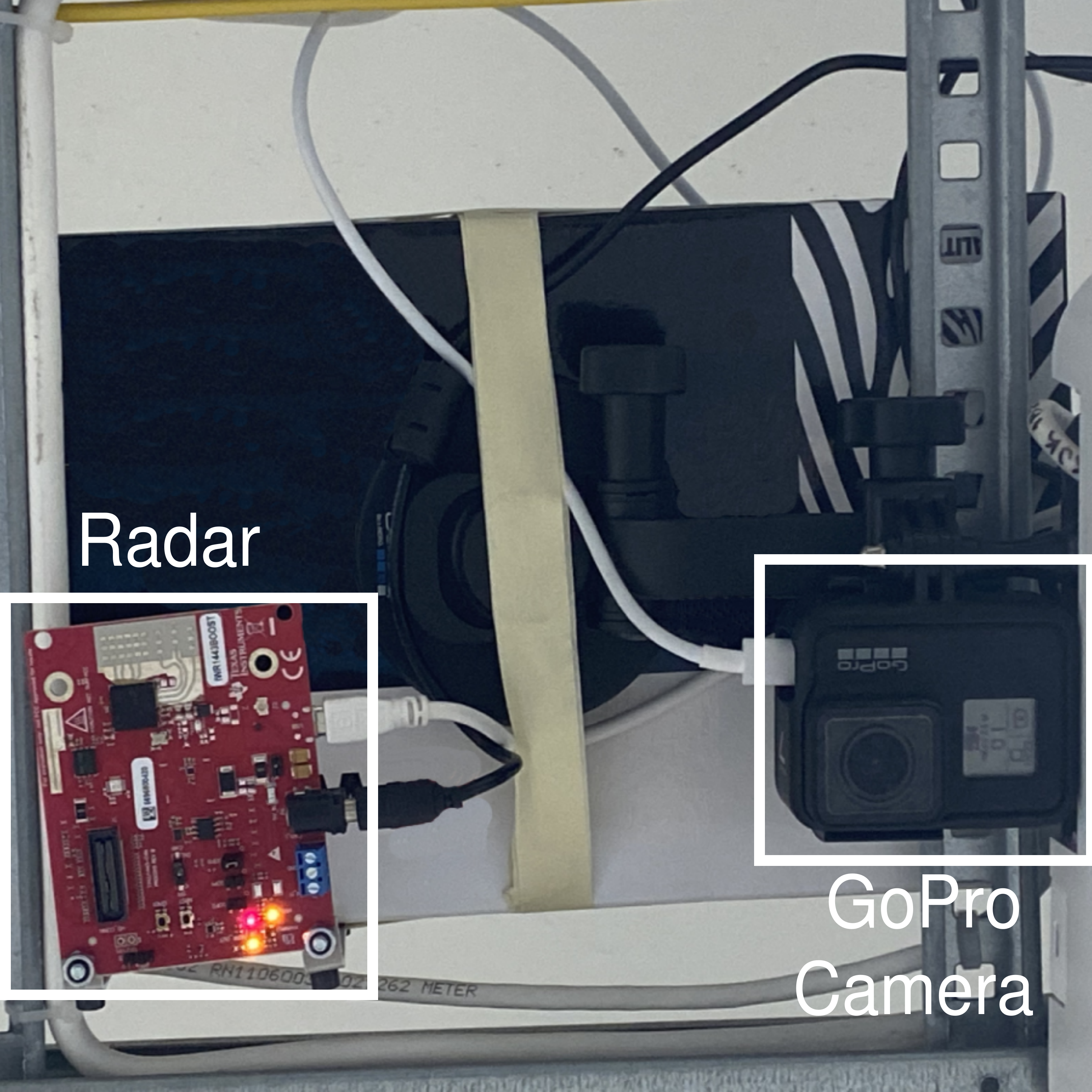}
  \caption{}
  \label{fig:radar_ceiling}
\end{subfigure}
\caption{(a) The experiment environment which is a circle with 2.5m radius marked on the ground. (b) An IWR1443 radar and a GoPro camera installed on the ceiling with 3m height.}
\label{fig:test}
\end{figure}

To collect the radar data (cf. Fig.~\ref{fig:radar_ground}), we marked a circular area with 2.5m radius on the floor and asked participants to walk arbitrarily inside the circle. 
A IWR1443 radar and a GoPro camera were mounted in 3m height over the center of the circle record ground truth point cloud and video data (Fig.~\ref{fig:radar_ceiling}). 
We labelled the video using the Ground Truth Labeler application. 
The radar data was labeled manually by matching the bounding boxes from the video with the point cloud data. 
Three subjects participated in the experiment and 15,000 samples have been collected. 
Training, validation and testing is performed following a 60/10/30 split.

For the \gls{lis}, we simulated the same circular area with 2.5m radius and simulated random positions of the target. 
The positions are processed by obtaining radio maps using $D_a=5$ active devices randomly deployed in the room. 
15,000 samples were recorded.

\subsection{Radar Model Training Process}
We used a computer with 32GB of RAM and an Nvidia GeForce 940MX GPU for training and inference. 
The model for the radar pipeline is implemented based on PyTorch and PyTorch Geometric~\cite{fey2019fast} frameworks.
An early stopping scheme with a patience of 100 epochs is exploited to prevent from over-fitting. 
In other words, if no improvement is observed on the validation set within the patience period, we stop the training process and save the best model.

\gls{mse} is used as the loss function to train the model, Adam Optimizer~\cite{kingma2014adam} with step-decay strategy to train the network:
\begin{equation}
    L_{r} = L_{i} \cdot d_{r} ^ {\lfloor \frac{e} { e_{r}}\rfloor}.
    \label{eq:step_decay}
\end{equation}
Here, $L_{r}$ is the learning, $L_{i}$ is the initial value of the learning rate, $d_{r}$ is the drop rate after every $e_{r}$ epochs, $e$ is the current epoch and $\lfloor\cdot\rfloor$ is the floor operator. 
In our experiments $L_{i}$, $d_{r}$, and $e_{r}$ are $0.001$, $0.5$, and $20$, respectively.

\subsection{LIS Radio Map Based Localization}
We process the radio map by obtaining the indices of the maximum peak value to infer the human position, i.e.

\begin{equation}
    (x_c, y_c) = \arg \max_{xp, yp}  |\mathbf{y_{mf}}| \times \Delta s,
\end{equation}
In this equation, $x_c$ and $y_c$ denote the inferred position, $x_p$ and $y_p$ the position in the pixel domain and $\Delta s$ the antenna spacing.

\subsection{LIS Bounding Box Creation}
As we are not using a learning algorithm, but want to compare \gls{lis} and radar systems, we compute manually a bounding box considering the center the inferred positions.

As we know the dimensions of our target (detailed in Section~\ref{subsubsec:Scenario}) we can compute a bounding box from the center position such that 
\begin{align}
    (x_{min}, y_{min}) = (x_c - \frac{L}{2}, y_c - \frac{W}{2}), \\
    (x_{max}, y_{max}) = (x_c + \frac{L}{2}, y_c + \frac{W}{2}),
\end{align}
with $L=0.5$ m and $W=0.3$ m (cf. Section~\ref{subsubsec:Scenario}).

\subsection{Evaluation Metric}
To evaluate the performance of the models we use \gls{iou} and the Euclidean distance between the center of the ground truth bounding box and the predicted bounding box. 
The \gls{iou} is defined by:

\begin{equation}
    \gls{iou}_i = \dfrac{|P_i \cap G_i|}{|P_i \cup G_i|},
\end{equation}

\noindent where $P_i$ and $G_i$ are predicted and ground truth bounding boxes for sample $i$.
We report the average \gls{iou} and Euclidean distance between the centers of the ground truth and the predicted bounding boxes on the test set for each of the approaches.

\subsection{Localization Comparison}
\begin{figure}
    \centering
    \includegraphics[width=1\columnwidth]{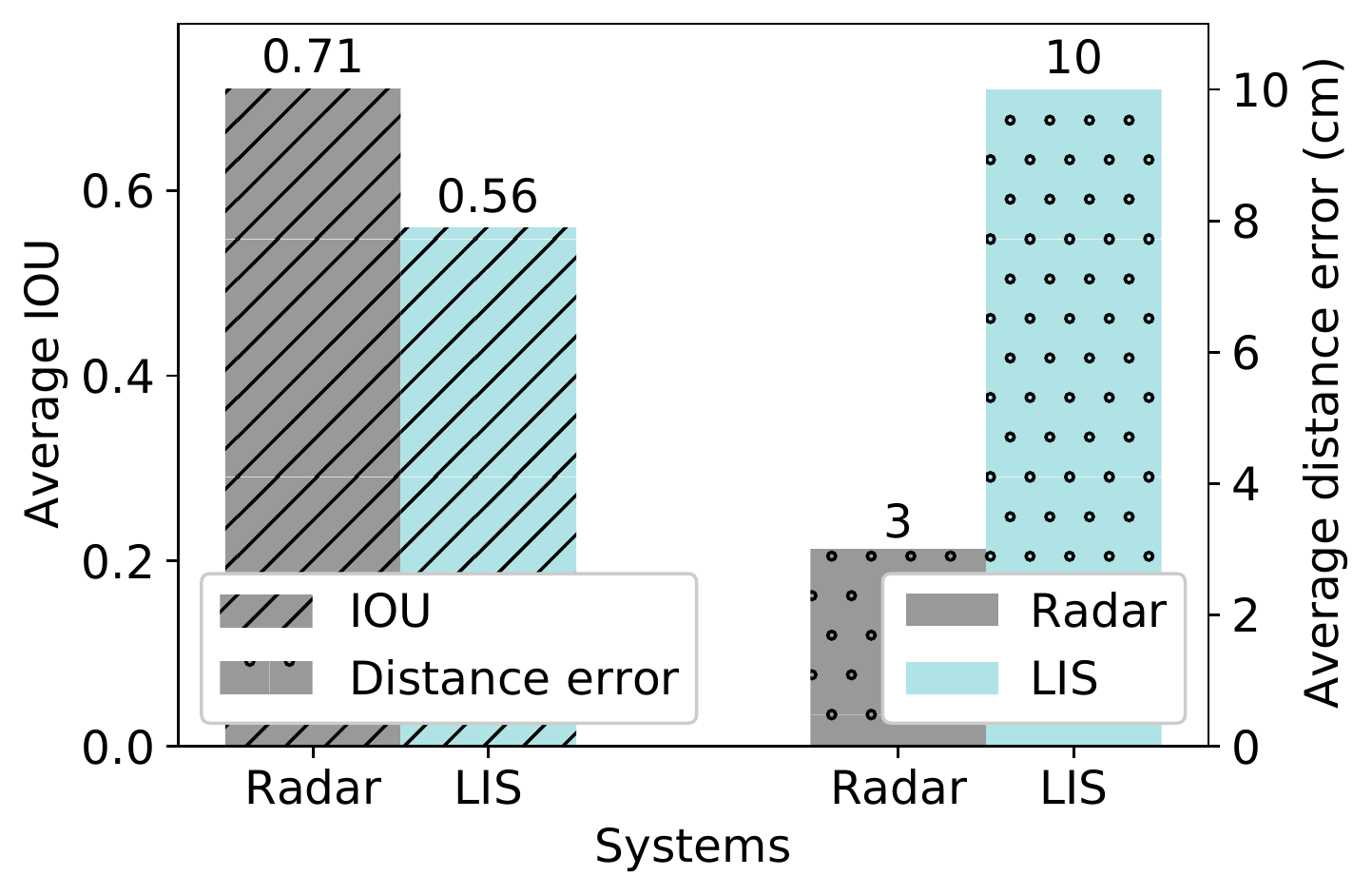}
    \caption{Average \gls{iou} and distance error for radar and \gls{lis}}
   \label{fig:results}
\end{figure}
As shown in Fig.~\ref{fig:results}, the average \gls{iou} for the \gls{lis} in localization is 0.56 while the Radar it is 0.71. 
This result suggests that the radar is more accurate in estimating the bounding box for the target. 
We have a similar result for the average distance error of the bounding boxes. 
The radar outperforms the \gls{lis} with only 3cm in contrast to 10cm average error. 
These results suggest that mmWave radars are more accurate than \gls{lis}, but in scenarios where the accuracy of sensing can be compromised, \gls{lis} can provide simultaneous communication and sensing capabilities without a need for a third-party device. 
Also, there might be potential scenarios in which \gls{lis} would outperform the radar. 
For example, decreasing the inter-antenna spacing would lead to more accurate localization results for the \gls{lis}. 
However, we showed the common inter-antenna distance design ($\frac{\lambda}{2}$) to provide a close comparison to the radar.
\section{Conclusion}
So far, mmWave radars have been considered the standard de facto for indoor localization given their incredibly cheap components and localization accuracy. 
However, the breakthroughs in communications systems are leading to devices that can integrate both sensing and communications. 
\glspl{lis} arise as a brand-new device which communications capabilities are expected to go beyond massive \gls{mimo}. 
However, their sensing functionality is still under investigation. 
In this paper, we have shown a comparison among \gls{lis} and mmWave radars in an indoor sensing-based localization task. 
Results show that \gls{lis} is not that different from mmWave radars in localization accuracy, with the advantage of leading to a new system that can integrate both sensing and communications. 
In most scenarios there is no need to have a very accurate localization, including but not limited to industrial localization systems or localizing users for beamforming. In those scenarios, \gls{lis} can also act as a sensing device providing communications and sensing capabilities at the same time.

\bibliographystyle{IEEEtran}
\bibliography{biblio}

\end{document}